# Steady State and Modulated Heat Conduction in Layered Systems Predicted by the Phonon Boltzmann Transport Equation

Jose Ordonez-Miranda[1,2], Ronggui Yang[2,*], Sebastian Volz[1], and J. J. Alvarado-Gil[3]

[1] Laboratoire d'Énergétique Moléculaire et Macroscopique, Combustion, UPR CNRS 288, Ecole Centrale Paris, Grande Voie des Vignes, 92295 Châtenay Malabry, France.

[2] Department of Mechanical Engineering, University of Colorado, Boulder, Colorado 80309, USA.

[3] Departamento de Física Aplicada, Centro de Investigación y de Estudios Avanzados del I.P.N-Unidad Mérida, Carretera Antigua a Progreso km. 6, A.P. 73 Cordemex, Mérida, Yucatán, 97310, México.

## Abstract

Based on the phonon Boltzmann transport equation under the relaxation time approximation, analytical expressions for the temperature profiles of both steady state and modulated heat conduction inside a thin film deposited on a substrate are derived and analyzed. It is shown that both steady state and modulated components of the temperature depend strongly on the ratio between the film thickness and the average phonon mean free path, and they exhibit the diffusive behavior as predicted by the Fourier's law of heat conduction when this ratio is much larger than the unity. In contrast, in the ballistic regime when this ratio is comparable to or smaller than the unity, the steady-state temperature tends to be independent of position, while the amplitude and the phase of the modulated temperature appear to be lower than those determined by the Fourier's law. Furthermore, we derived an invariant of heat conduction and a simple formula for the cross-plane thermal conductivity of dielectric thin films, which could be a useful guide for understanding and optimizing the thermal performance of the layered systems. This work represents the Boltzmann transport equation-based extension of the Rosencwaig and Gerko work [J. Appl. Phys. **47,** 64 (1976)], which is based on the Fourier's law and has widely been used as the theoretical framework for the development of photoacoustic and photothermal techniques.

---

[*] **Corresponding author**: Telephone: +1 303 735 1003 and Email: ronggui.yang@colorado.edu

This work might shed some light on developing a theoretical basis for the determination of the phonon MFP and relaxation time using ultrafast laser-based transient heating techniques.

## I. INTRODUCTION

Thermal transport across solid dielectric films has been widely studied, due to their wide range of applications, especially in nanoelectronics, optoelectronics, and micro/nano-electromechanical systems (M/NEMS).[1-3] The thickness of these dielectric films can be comparable to or smaller than the mean free path (MFP) of the heat carriers, *i.e.,* phonons or electrons, and therefore the conventional Fourier's law of heat conduction can no longer be applied.[4-6] The Fourier's law is valid in macroscopic systems where phonons have enough space to interact each other and reach their local thermodynamic equilibrium.[7,8] Figure 1 shows three different scenarios of heat conduction across a film of thickness $L$ with temperatures $T_1$ and $T_2$ at their boundaries ($T_1 > T_2$). In the diffusive regime, for which the film thickness is much larger than the phonon MFP ($L \gg l$), the Fourier's law is valid since there are sufficient phonon-phonon collisions which guarantee the establishment of a temperature gradient (Fig. 1(a)). When the film thickness is much smaller than the phonon MFP ($L \ll l$), the phonon-phonon interactions inside the film is rare and the temperature concept cannot be established due to the size requirement and hence the Fourier's law cannot be applied without a temperature gradient, as shown in Fig. 1(b). In the pure ballistic limit, the conduction process is in analogue to the thermal radiation and the heat flux across the film is given by $q = \sigma(T_1^4 - T_2^4)$,[9,10] where $\sigma$ is the analogous Stephan-Boltzmann constant for phonons. Taking into account that the phonon MFP is from a few nanometers to tens of microns for a wide variety of dielectric materials at room temperature,[8,10] the heat conduction across thin films are likely to be determined by a combination of the diffusive and ballistic transport, as shown in Fig. 1(c). In the meantime, the switching transient of micro/nano-devices is readily approaching the phonon relaxation time, which also invalidates the prediction of the transient thermal conduction process using the Fourier's law.

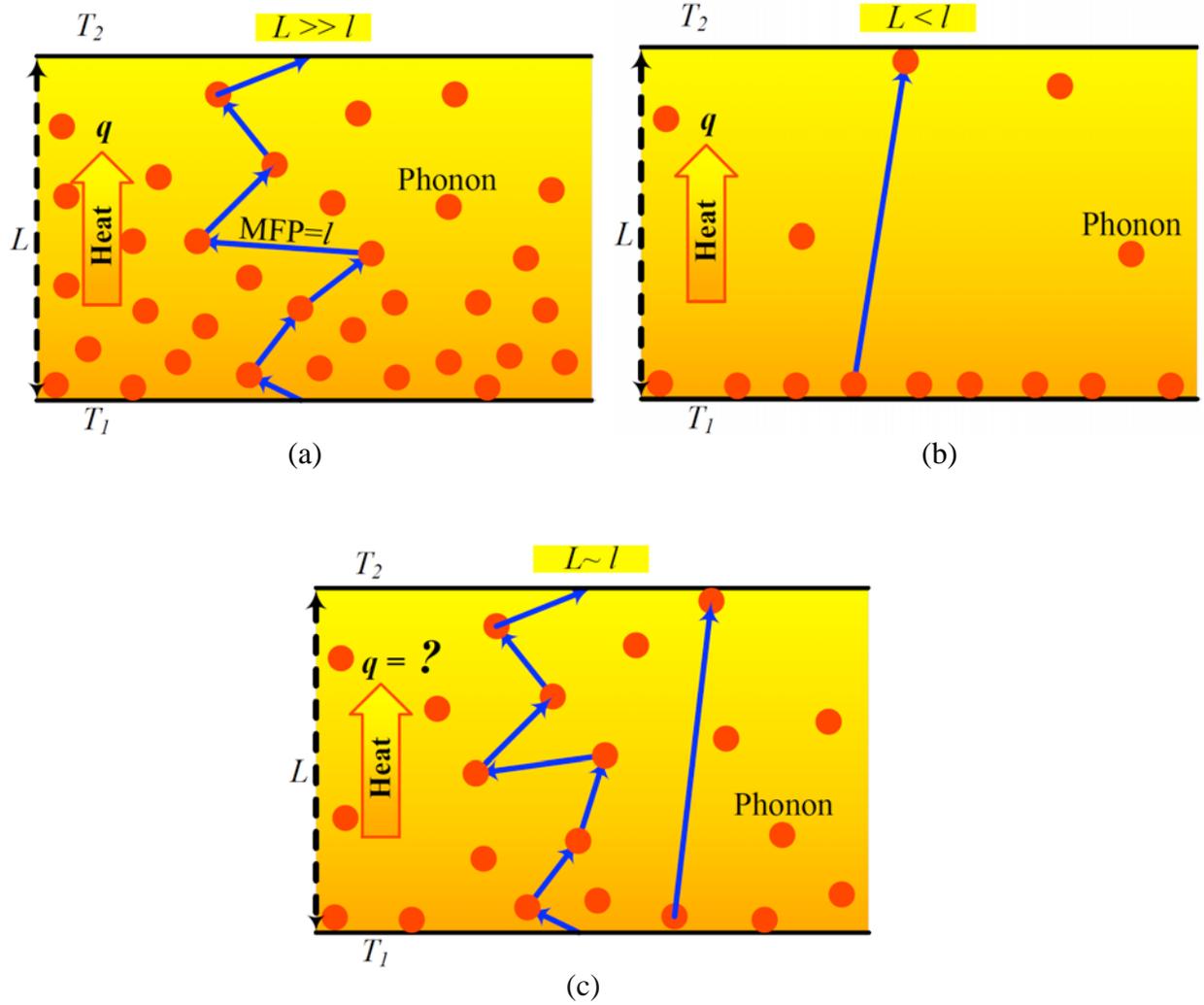

*FIG. 1. Phonon dynamics in the (a) diffusive, (b) ballistic, and (c) quasi-ballistic (intermediate) regimes of heat conduction, depending on the ratio of thin film thickness and phonon mean free path.*

It has been shown that the phonon Boltzmann transport equation (BTE) is a more appropriate tool to describe the transport phenomena in nanostructured materials and during the ultrafast processes.[4,6] Though great progresses have been made in solving the BTE for micro/nanoscale heat conduction with significant efforts in recent years, the inherent difficulties associated with its solution have significantly limited the consideration of the size and transient effects. Under the steady-state conditions, the BTE has been solved numerically and applied to study the heat transport through a variety of layered systems and complex geometries.[1,4,5,11,12] These works showed the two main findings: i) a reduction of the effective thermal conductivity with respect to their bulk values, and ii) the temperature profiles differ significantly from those obtained using the Fourier's law, due to the ballistic behavior of the energy carriers. The transient heat conduction in thin films has also been revisited in some recent works using the transient phonon BTE.[13-15] Taking into account that the energy carriers travel ballistically without being deflected out of their propagation direction in a spatial scale in the order of one MFP, Chen proposed the ballistic-diffusive equations to study transient heat conduction from macro- to nanoscales.[13,14] Even though this model presents good agreements with the predictions of the BTE for the heat conduction, it cannot be easily implemented. More recently Ordonez-Miranda et al.[15] proposed the multipole model of heat conduction based on the solution of the BTE. Their results agree well with the phonon BTE, but the closed form solution of the model can be obtained analytically only in a few limiting cases. Approximate methods based on the BTE, that are capable of capturing the size affects, but easy to implement analytically or numerically are thus still desired.[16]

On the other hand, photothermal measurements, including both frequency-domain and time-domain transient thermoreflectance techniques using modulated ultrafast laser heating, have been widely utilized for characterizing thermal properties of materials. In transient thermoreflectance measurement, an ultrafast laser beam with a sub-pico second pulse width is often modulated with a frequency of several hundreds Hz to tens of mega Hz (MHz) to characterize thermal properties of thin films and interface thermal resistance. At this length and time scale, the Fourier's law of heat conduction might not be strictly valid. However, the solutions of the Fourier's law of heat conduction is often used to fit the experimental measurement to obtain the effective thermal properties due to the lack of accessible solution of the BTE.[17-19] More recently, both the frequency-domain and time-domain thermoreflectance methods have been utilized to extract the mean free path and the relaxation time of phonons, as fundamental properties of materials after the inspiring work by Siemens et al.[20-25] Clearly there is a strong need of phonon BTE solutions to better understand and validate data reduction schemes in these experiments which at present are dominantly through fitting an effective thermal conductivity or conductance with the Fourier's law.[18]

The objective of this work is to solve analytically the phonon Boltzmann transport equation for the temperature profile in a dielectric film deposited on a substrate, when excited with a modulated laser beam. [text obscured by figure overlays referring to $\tau$] strongly on relative thickness [obscured] phonon MFP and they agree quite well with the numerical results previously reported in the literature.[1,6] The amplitude and the phase of the modulated temperature, on the other hand, depend strongly on the product of the modulation frequency [obscured] time, such that they reduces to the corresponding [obscured] when this product [obscured] work represents [obscured] ork by Rosencwaig [obscured] based on the Fourier's law of heat conduction and has widely been used as the theoretical basis for the development of photoacoustic and photothermal techniques. This work might shed some light on developing sound theoretical basis for the determi[nation of] relaxation time using ultrafast laser-based transient thermorefle[ctance obscured].

## II. PHONON HEAT CONDUCTION IN A TWO-LAYER SYSTEM

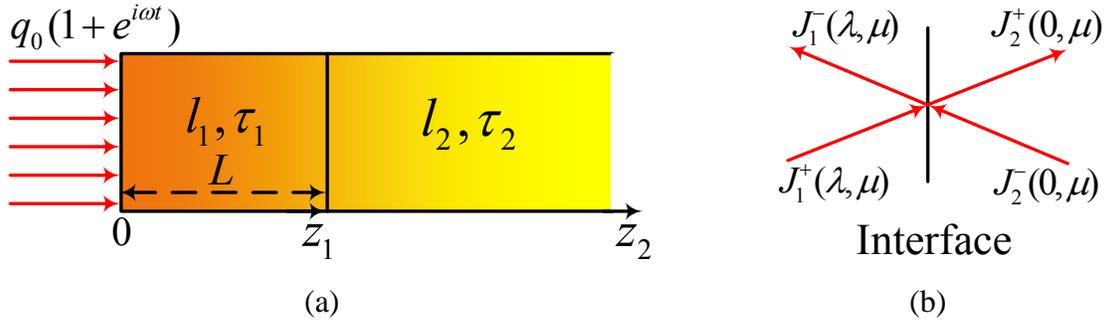

(a)          (b)

*FIG. 2. Schematics of (a) a two-layer system under consideration, and (b) the transmitted and reflected phonon intensities at the interface $z_1 = L$ ($z_2 = 0$).*

Let us now consider a two-layer system shown in Fig. 1(a), where a dielectric film with a finite thickness is heated by a laser beam with modulated intensity, at the surface $z_1 = 0$. This system represents a thin film deposited on a substrate in analogue of sample configuration for photo-thermal characterization of thermal conductivity. The surface heat flux $q$ at $z_1 = 0$ by the external thermal excitation is given by[26]

$$q(z_1 = 0, t) = q_0 \left(1 + \cos(\omega t)\right) = q_0 \operatorname{Re}\left(1 + e^{i\omega t}\right), \quad (1)$$

where $\omega$ is the angular modulation frequency, $2q_0$ is the average intensity of the laser, $t$ is the time and Re() stands for the real part of its argument. The steady state and modulated temperature fields within the layers, for the surface heat flux defined in Eq.(1) can be easily found using the Fourier's law and other revised Fourier-like models of heat conduction.[2,27,28] However, considering that the MFP $l_1$ of the energy carriers inside the film can be comparable to or even smaller than the thickness $L$ ($l_1 < L$), these macroscopic models might not be valid.[6] We hereby develop analytical solutions using the transient phonon BTE.

For the sake of simplicity and to present the results analytically, this work will be developed assuming that the scattering processes inside the layers are described by the average values of the MFP ($l$) and relaxation time ($\tau$) of phonons, such that these parameters are independent of the phonon frequency. Assuming that the changes of temperature due to heating is small, the phonon BTE, under the single mode relaxation time approximation and in the intensity representation, can then be written as[1]

$$I + \tau \frac{\partial I}{\partial t} + \mu l \frac{\partial I}{\partial z} = I_0, \tag{2}$$

where $\mu = \cos(\theta)$ is the cosine of the angle between the phonon propagation direction and the $+z$ axis, $I$ is the total phonon intensity, and $I_0$ is the "equilibrium" phonon intensity defined by[1,5]

$$I_0 = \frac{1}{4\pi} \int_0^{\upsilon_{max}} h\upsilon v f_0 D(\upsilon) d\upsilon, \tag{3}$$

being $h$ the Planck constant, $\upsilon$ the phonon frequency, $v$ the magnitude of the phonon group velocity, $f_0$ the Bose-Einstein equilibrium distribution function, and $D$ the density of states per unit volume. The phonon intensity $I$ is related to the phonon distribution function $f$ through Eq. (3), with the interchanges $I_0 \to I$ and $f_0 \to f$. Under the Debye model,[6] $I_0$ can be written as

$$I_0 = \frac{15\sigma T^4}{\pi^5} \int_0^{T_D/T} \frac{x^3 dx}{e^x - 1}, \tag{4}$$

where $\sigma = \pi^5 k_B^4 / (5h^3 v^2)$ is the analogous Stephan-Boltzmann constant for phonons, $k_B$ is the Boltzmann constant, $T$ is the temperature and $T_D$ is the Debye temperature. At low temperature ($T << T_D$), Eq.(4) reduces to

$$I_0 = \frac{\sigma}{\pi} T^4, \qquad (5)$$

which is analogous to the well-known Stephan-Boltzmann law for photons. On the other hand, at high temperature ($T >> T_D$), Eq.(4) takes the form

$$I_0 = \frac{15\sigma T_D^3}{3\pi^5} T = \frac{\rho c v}{4\pi} T, \qquad (6)$$

where $\rho$ is the mass density and $c$ the specific heat at high temperature.[6] Taking into account the definition and the temperature dependence of the specific heat $c$,[8,10] Eqs. (3) and (5) shows that $I_0$ can be written in the form of the second equality in Eq. (6), i.e., linear relationship between the equilibrium intensity and temperature, for both low and high temperatures. We will use this relationship that the equilibrium intensity is proportional to the temperature to express some results of the present work.

If the intensity $I$ in the one-dimensional phonon BTE is solved, the heat flux $q$ can then be determined as follows[15]

$$q(z,t) = 2\pi \int_{-1}^{1} \mu I(z,t,\mu) d\mu. \qquad (7)$$

For the surface thermal excitation defined in Eq.(1), the solution for the intensity $I$ of Eq. (2) has the following general form

$$I(z,t,\mu) = I_s(z,\mu) + \text{Re}\left[J(z,\mu)e^{i\omega t}\right], \qquad (8)$$

where $I_s$ and $J$ are the stationary and modulated components of $I$, due to the first and second terms of the external thermal excitation, respectively. Given that Eq. (2) is a linear partial differential equation, Eq.(8) implies that the equilibrium intensity $I_0$ should have the same time dependence as the total intensity $I$. Therefore, the general form of $I_0$ is

$$I_0(z,t) = I_{0s}(z) + \text{Re}\left[J_0(z)e^{i\omega t}\right]. \qquad (9)$$

Furthermore, given the direct proportionality between the equilibrium intensity and temperature for small temperature changes as articulated earlier, Eqs. (6) and (9) establish that $T(z,t) \to T(z) + \text{Re}\left[\psi(z)e^{i\omega t}\right]$, where $T(z)$ and $\psi(x)$ are the steady state and modulated components of temperature. After inserting Eqs. (8) and (9) into Eq.(2), the following two uncoupled differential equations can be written for $I_s$ and $J$, respectively:

$$I_s + \mu \frac{\partial I_s}{\partial x} = I_{0s}, \tag{10a}$$

$$\chi J + \mu \frac{\partial J}{\partial x} = J_0, \tag{10b}$$

where $x = z/l$ and $\chi = 1 + i\omega\tau$. Note that the frequency-dependence of $J$ comes through the complex parameter $\chi$. When $\omega = 0$, Eq.(10b) reduces to its stationary counterpart Eq. (10a). Given that the typical values of the average relaxation time are $\tau < 10^{-10}$ s for a wide variety of materials at room temperature,[6] the parameter $\chi \to 1$, for any periodic heating with frequency $f = \omega/2\pi << 1\text{ GHz}$. This value indeed covers almost all the operating frequency range of existing photo-thermal and photoacoustic techniques. For these reasons, we are going to derive the solutions for the modulated heat transport with the assumption $\chi \leq 1$. The solution of Eq. (10b) is straightforward and is given by

$$J(x,\mu) = J(x_0,\mu)e^{\chi(x_0-x)/\mu} + \frac{1}{\mu}\int_{\xi_1}^{\xi} J_0(x')e^{\chi(x'-x)/\mu} dx', \tag{11}$$

where $x_0$ is an integration constant. Due to the interface roughness,[6] usually diffuse interface scattering dominates and we will limit our discussion to diffuse interface scattering in this work. For diffuse surface/interface scattering, the intensities leaving the surfaces at $z = 0, L$ are uniform,[11] and therefore the coefficients $J(x_0,\mu)$ should also be independent of the direction $\mu$. To facilitate the evaluation of the boundary conditions, it is convenient to split Eq. (11) into two parts, one for each layer of the system shown in Fig. 1(a), as follows[5]

$$J_1^+(x,\mu) = \frac{A_1^+}{\pi}e^{-\chi_1 x/\mu} + \frac{1}{\mu}\int_0^x J_0(x')e^{\chi_1(x'-x)/\mu} dx', \tag{12a}$$

$$J_1^-(x,\mu) = \frac{A_1^-}{\pi} e^{\chi_1(\lambda-x)/\mu} - \frac{1}{\mu}\int_x^\lambda J_{01}(x')e^{\chi_1(x'-x)/\mu}\,dx', \tag{12b}$$

$$J_2^+(x,\mu) = \frac{A_2^+}{\pi} e^{-\chi_2 x/\mu} + \frac{1}{\mu}\int_0^x J_{02}(x')e^{\chi_2(x'-x)/\mu}\,dx', \tag{12c}$$

$$J_2^-(x,\mu) = -\frac{1}{\mu}\int_x^\infty J_{02}(x')e^{\chi_0(x'-x)/\mu}\,dx', \tag{12d}$$

where $\lambda = L/l_1$, $\chi_n = 1 + i\omega\tau_n$, the superscript $+$ ($-$) stands for $0 \leq \mu \leq 1$ ($-1 \leq \mu \leq 0$), as shown in Fig. 2(b), the constants $A_1^\pm$ and $A_2^+$ are determined by the boundary conditions, and the subscripts $n=1$ and $n=2$ stand for the first and second layers, respectively. The temperature field is determined by the principle of energy conservation, which in terms of the phonon intensities $I_0$, can be written as[1]

$$2I_0 = \int_{-1}^1 I\,d\mu. \tag{13}$$

For the modulated components of the intensities $J_0$ and $J$, Eq. (13) yields

$$2J_0(x) = \int_0^1 \left[J^+(x,\mu) + J^-(x,-\mu)\right]d\mu, \tag{14}$$

After inserting Eqs. (12a)-(12d) into Eq.(14), the following equations for $J_0$ inside the first ($J_{01}$) and second ($J_{02}$) layers are obtained

$$2J_{01}(x) = \frac{A_1^+}{\pi}E_2(\chi_1 x) + \frac{A_1^-}{\pi}E_2(\chi_1(\lambda-x)) + \int_0^\lambda J_{01}(x')E_1(\chi_1|x-x'|)\,dx' \tag{15a}$$

$$2J_{02}(x) = \frac{A_2^+}{\pi}E_2(\chi_2 x) + \int_0^\infty J_{02}(x')E_1(\chi_2|x-x'|)\,dx' \tag{15b}$$

where $E_n()$ is the exponential integral function of order $n$.[29] Based on Eqs. (7) and (8), the modulated component $q_t$ of the heat flux in each layer can be written as

$$q_t(x,t) = 2\pi\int_{-1}^1 \left(J^+(x,\mu) - J^-(x,-\mu)\right)\mu\,d\mu. \tag{16}$$

The combination of Eqs. (12a)-(12d) and Eq. (16) yields the following modulated heat fluxes inside the first ($q_{t1}$) and second ($q_{t2}$) layers

$$\frac{q_{t1}(x)}{2} = A_1^+ E_3(\chi_1 x) - A_1^- E_3(\chi_1(\lambda - x)) - \frac{\pi}{\chi_1}\frac{d}{dx}\int_0^\lambda J_{01}(x')E_3(\chi_1|x-x'|)dx', \tag{17a}$$

$$\frac{q_{t2}(x)}{2} = A_2^+ E_3(\chi_2 x) - \frac{\pi}{\chi_2}\frac{d}{dx}\int_0^\infty J_{02}(x')E_3(\chi_2|x-x'|)dx'. \tag{17b}$$

The constant $A_1^\pm$ and $A_2^+$ can be determined by imposing the energy balance at the interfaces of the layers. According to Eq. (7) and Fig. 2(b), the conservation of energy at the inner interface of the layers establishes that

$$\int_0^1 J_1^-(\lambda,-\mu)\mu d\mu = r_{12}\int_0^1 J_1^+(\lambda,\mu)\mu d\mu + t_{21}\int_0^1 J_2^-(0,-\mu)\mu d\mu, \tag{18b}$$

$$\int_0^1 J_2^+(0,\mu)\mu d\mu = r_{21}\int_0^1 J_2^-(0,-\mu)\mu d\mu + t_{12}\int_0^1 J_1^+(\lambda,\mu)\mu d\mu, \tag{18b}$$

where $r_{ij}$ and $t_{ij}$ are the energy reflectivity and transmissivity of phonons coming from the $i^{th}$ layer toward the $j^{th}$ layer, respectively. Considering that under the diffuse scattering, the scattered phonons completely lose their memory,[5] these coefficients are direction-independent and satisfy the relations[5,11]

$$t_{ij} = r_{ji} = 1 - t_{ji} = \frac{\rho_j c_j v_j}{\rho_i c_i v_i + \rho_j c_j v_j}. \tag{19}$$

On the other hand, the energy balance at the illuminated surface $z = 0$ yields

$$2\pi\int_0^1 J_1^+(0,\mu)\mu d\mu = 2\pi\int_0^1 J_1^-(0,-\mu)\mu d\mu + q_0, \tag{20}$$

which is a particular case of Eq. (16). The combination of Eqs. (12), (18), and (20) yields the following system of equations for the constants $A_1^\pm$ and $A_2^+$

$$\frac{A_2^+}{t_{12}} = \frac{A_1^-}{r_{12}}, \tag{21a}$$

$$\frac{A_2^+}{2t_{12}} = A_1^+ E_3(\chi_1\lambda) + \int_0^\lambda \pi J_{01}(x) E_2(\chi_1(\lambda-x)) dx + \int_0^\infty \pi J_{02}(x) E_2(\chi_2 x) dx, \qquad (21b)$$

$$A_1^+ = q_0 + 2A_1^- E_3(\chi_1\lambda) + 2\int_0^\lambda \pi J_{01}(x) E_2(\chi_1 x) dx. \qquad (21c)$$

Under the boundary conditions in Eqs. (21a)-(21c), Eqs. (15) and (17) along with Eq. (4), determine fully the modulated heat transfer through a finite layer on a semi-infinite substrate. The steady-state component of the equilibrium intensity $I_{0s}$ and the heat flux $q_s$ can be easily determined by replacing $J_{0n} \to I_{0sn}$, $q_{tn} \to q_{sn}$, and $\chi_n = 1$ in Eqs. (15), (17), and (21). Below we analyze separately the steady state and modulated heat conduction problems.

## A. Steady-state heat conduction

Under steady state condition, Eqs. (15) and (17) can be normalized as follows

$$2U_1(x) = E_2(x) + \int_0^\lambda U_1(x') E_1(|x-x'|) dx' \qquad (22a)$$

$$2U_2(x) = E_2(x) + \int_0^\infty U_2(x') E_1(|x-x'|) dx' \qquad (22b)$$

$$\frac{Q_{s1}}{2} = E_3(x) - \frac{d}{dx}\int_0^\lambda U_1(x') E_3(|x-x'|) dx', \qquad (22c)$$

$$\frac{Q_{s2}}{2} = E_3(x) - \frac{d}{dx}\int_0^\infty U_2(x') E_3(|x-x'|) dx', \qquad (22d)$$

where

$$U_n(x) = \frac{\pi I_{0sn}(x) - A_n^-}{A_n^+ - A_n^-}, \qquad (23a)$$

$$Q_{sn}(x) = \frac{q_{sn}(x)}{A_n^+ - A_n^-}, \qquad (23b)$$

with $A_2^- = 0$. By taking the derivative of Eqs. (22c) and (22d) and comparing the results with the corresponding equations (22a) and (22b), it can be shown that $Q_{s1}$ and $Q_{s2}$ are constants. This is consistent with the principle of energy conservation, which establishes that $q_{s1} = q_{s2} = q_0$. In what follows, we are going to obtain and discuss the analytical solutions of Eqs. (22a) and (22b).

- **Semi-infinite layer**

Based on the properties of the function $E_n()$,[29] it can be seen that the unity is a particular solution of Eq. (22b). Therefore the general solution for Eq. (22b) can be written as $U_2(x) = 1 - G(x)$. In terms of the function $G(x)$, Eqs. (22b) and (22d) reduce to

$$2G(x) = \int_0^\infty G(x')E_1(|x-x'|)dx', \qquad (24a)$$

$$\frac{Q_{s2}}{2} = \int_0^\infty G(x+x')E_2(x')dx' - \int_0^x G(x-x')E_2(x')dx'. \qquad (24b)$$

Given the direct proportionality between the equilibrium intensity and temperature, from the Fourier's law of heat conduction, we know that in the diffusive limit ($x \to \infty$), $G(x) = a + bx$, where $a$ and $b$ are constants. In this limit, we insert the function $G$ in Eq. (20b) and evaluate explicitly the involved integrals to obtain $b = 3Q_{s2}/4$. Therefore, the general solution of Eq. (24a) can be written as

$$G(x) = \frac{3}{4}Q_{s2}\bigl(x + A + g(x)\bigr), \qquad (25)$$

where $A$ is a constant and the function $g(x) \to 0$ for $x \to \infty$. The combination of Eqs. (24a) and (25) yields the following integral equation for $g(x)$

$$2g(x) = -AE_2(x) + E_3(x) + \int_0^\infty g(x')E_1(|x-x'|)dx', \qquad (26a)$$

which suggests that for a first-order approximation, $g(x)$ is given by

$$g(x) = BE_2(x) + CE_3(x). \qquad (26b)$$

The constants $A$, $B$, and $C$ can be calculated by evaluating Eqs. (24a) and (24b) at $x = 0$:

$$2G(0) = \int_0^\infty G(y)E_1(y)dy, \qquad (27a)$$

$$\frac{Q_{s2}}{2} = \int_0^\infty G(y) E_2(y) dy. \tag{27b}$$

The required third equation can be obtained by expressing the right-hand side of Eq. (24b) as an exact derivative and integrating on both sides. The final result is:

$$\frac{Q_{s2}}{2}(x+\delta) = \int_0^\infty G(x') E_3(|x-x'|) dx'. \tag{28}$$

where $\delta$ is an integration constant. The evaluation of Eq. (28) in the diffusive limit ($x \to \infty$) yields $\delta = A$. Therefore, at $x = 0$, Eq. (28) reduces to

$$\frac{Q_{s2} A}{2} = \int_0^\infty G(y) E_3(y) dy. \tag{29}$$

By inserting Eqs. (25) and (26b) into Eqs. (27) and (29), the following system of equations for $A$, $B$, and $C$ is obtained

$$\begin{pmatrix} 1 & 2-I_{12} & 1-I_{13} \\ 1/2 & I_{22} & I_{23} \\ 1/3 & -I_{23} & -I_{33} \end{pmatrix} \begin{pmatrix} A \\ B \\ C \end{pmatrix} = \begin{pmatrix} 1/2 \\ 1/3 \\ 1/4 \end{pmatrix}, \tag{30}$$

where the coefficients $I_{nm}$ are defined by

$$I_{nm} = \int_0^\infty E_n(y) E_m(y) dy, \tag{31}$$

which can now be calculated analytically.[29]

The solution of Eq. (30) for five decimal figures is $A = 0.71047$, $B = -0.25082$, and $C = 0.23526$. Hence, the approximate solution of Eq. (29a) can be written as

$$G(x) = \frac{3}{4} Q_{s2} (x + p(x)), \tag{32a}$$

$$p(x) = A + B E_2(x) + C E_3(x). \tag{32b}$$

By comparing Eq. (32a) with the solution for a semi-infinite heat conduction problem using the Fourier's law, it is clear that the non-Fourier (ballistic) contribution to the temperature is contained in $p(x)$, which is a positive function that increases monotonously with $x = z/l$, as shown in Fig. 3. The minimum and maximum values for $p(x)$ are $p(0) = 1/\sqrt{3}$ and $p(\infty) = A$, respectively. The first of these values has been obtained using the exact values of $A$, $B$, and $C$. A second-order approximation for $f(x)$ can directly be obtained by inserting Eq. (32) into the right-hand side of Eq. (24a). The mathematical expression of this iterated solution is much longer than Eq. (32a), but in numbers they differs in less than 0.1% for any $x \geq 0$. This fast convergence for the solution of $f(x)$ is reasonable given that the exponential decay of the integral exponential functions $E_n(x)$ as $x$ increases. The fast convergence of expansions involving the functions $E_n(x)$ was also found and verified in radiative heat transfer.[30] Hence, for applications of practical interest, Eq. (32a) provides a good approximate solution for the temperature $U_2(x) = 1 - f(x)$ within a semi-infinite layer.

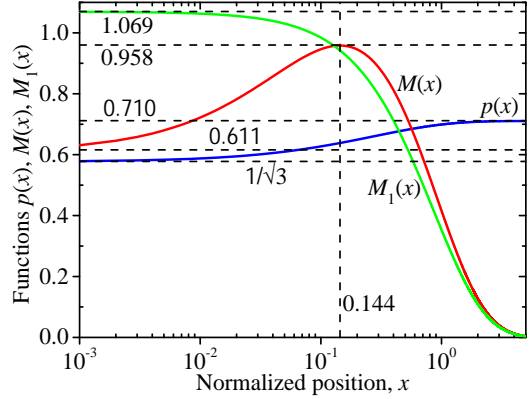

**FIG. 3**. *Characteristic functions p(x) and M(x) and $M_1(x)$ involved in the steady state and modulated heat conduction problems.*

In absence of the layer with finite thickness in Fig. 2(a), and considering that the temperature at the illuminated surface of the semi-infinite layer is $T_0$, Eqs. (6) and (32) establish that the explicit expression for the temperature field $T$ inside this semi-infinite layer is

$$T_0 - T(x) = \frac{3q_0}{\rho c v}\left(x + p(x)\right), \tag{33}$$

which differs from the Fourier's law prediction by the function $1/\sqrt{3} \leq p(x) \leq A \approx 1/\sqrt{2}$. For a position inside the semi-infinite layer with a distance much larger than the phonon MFP from the illuminated surface, i.e., $x = z/l \gg 1$, Eq. (33) reduces to the diffusive prediction, as should be.

- **Thin Film Layer**

Here we derive an explicit expression for the steady state temperature inside a thin film of thickness $L$. This will allows us comparing our analytical approach with previous numerical results reported for this system.[1,6] Based on Eq. (A19) of the appendix, the solution of Eq. (22a) for a thin film layer is given by

$$U_1(x) = 1 - \alpha\left[x + \beta + \gamma\left(p(x) - p(\lambda - x)\right)\right], \tag{34a}$$

$$\alpha = \frac{3}{4}Q_{s1} = \frac{1}{\lambda + 2\beta}. \tag{34b}$$

To find the parameters $\beta$ and $\gamma$, we evaluate Eqs. (22a) and (22c) at $x = 0$:

$$2U_1(0) = 1 + \int_0^\lambda U_1(y)E_1(y)dy, \tag{35a}$$

$$Q_{s1} = 1 - 2\int_0^\lambda U_1(y)E_2(y)dy. \tag{35b}$$

The combination of Eqs. (34) and (35) yields the following system of equations

$$\begin{pmatrix} 1 - E_2(\lambda) & -\xi \\ 1/2 + E_3(\lambda) & C_2 \end{pmatrix} \begin{pmatrix} \beta \\ \gamma \end{pmatrix} = \begin{pmatrix} 1/2 - E_3(\lambda) \\ 1/3 + E_4(\lambda) \end{pmatrix}, \tag{36}$$

where $\xi = C_1 + 2\left(p(\lambda) - p(0)\right)$ and

$$C_n = \int_0^\lambda \left(p(y) - p(\lambda - y)\right)E_n(y)dy. \tag{37}$$

Figure 4 shows the behavior of the parameters $\beta$, $\gamma$ and of the normalized heat flux $Q_{s1}$ as a function of the normalized layer thickness $\lambda = L/l_1$. Note that the values of these three parameters $\beta$, $\gamma$ $Q_{s1}$ are bounded, such that $\beta$ and $\gamma$ reach their maxima when $\lambda = L/l_1$ is close to 1. For a semi-infinite layer ($\lambda \to \infty$), $\beta \to p(\infty)$, $\gamma \to 1$, and Eq. (34a) reduces to Eq. (32a) of the semi-infinite layer. On the other hand, for a very thin layer with $\lambda << 1$, the normalized heat flux reduces to $Q_{s1} = 1$, which agrees with Eq. (35b). Assuming that the layer surfaces $x = 0, \lambda$ are at temperatures $T_1$ and $T_2(< T_1)$, then this last condition renders that the heat flux across the layer is $q_{s1} = \sigma\left(T_1^4 - T_2^4\right)$, where $\sigma$ is the analogous Stephan-Boltzmann constant for phonons as defined in Eq. (4). The results at the ballistic limit is very different from the prediction of Fourier's law in the diffusive regime and coincide with the result reported by Swartz and Pohl,[9] in absence of phonon scattering within a film. Furthermore, for an arbitrary $\lambda > 0$ and assuming that the temperature difference $T_1 - T_2$ is small enough, Eqs. (6) and (34b) establishes that the heat flux across the layer is given by $q_{s1} = k_0(1 + 2\beta/\lambda)^{-1}(T_1 - T_2)/L$, which has the form of the Fourier's law but with a modified thermal conductivity

$$k = \frac{k_0}{1 + 2\beta/\lambda}, \tag{38}$$

where $k_0 = \rho c v l/3$ is the bulk thermal conductivity of the layer. For $\lambda \ll 1$, $\beta \to 2/3$ and Eq. (35) reduces to the numerical result reported by Majumdar[1] and Chen.[6] Given that Eq. (38) has been rigorously derived from an analytical approach, this result represents an accurate extension of those previous results for the cross-plane thermal conductivity of a thin film.

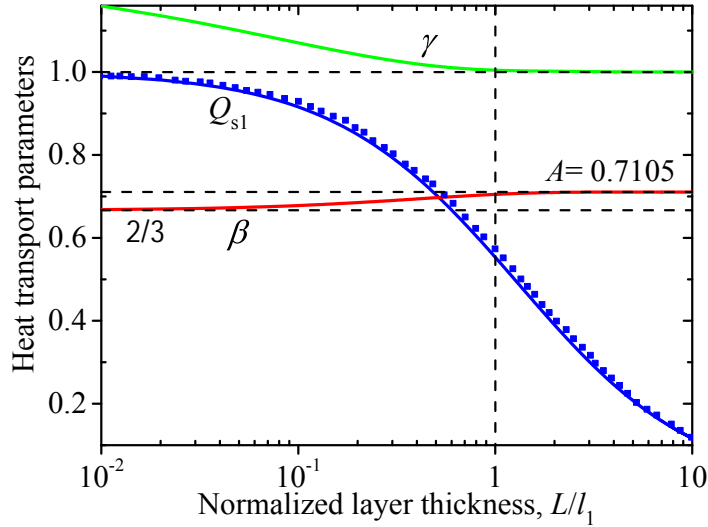

*FIG. 4. Heat transport parameters involved in the equilibrium intensity for the steady state temperature. The dashed line stands for the numerical predictions by Chen.[6]*

- **Two-layer system**

For the two-layer system shown in Fig. 2(a), according to Eq. (34a), the equilibrium intensity $I_{0s1}$ inside the finite layer can be written as

$$A_1^+ - \pi I_{0s1}(x) = \frac{3q_0}{4}\left[x + \beta + \gamma\left(p(x) - p(\lambda - x)\right)\right]. \tag{39a}$$

Furthermore, based on Eqs. (23a) and (32a), the intensity $I_{0s2}$ inside the semi-infinite layer is given by

$$A_2^+ - \pi I_{0s2}(x) = \frac{3q_0}{4}(x + p(x)), \tag{39b}$$

where the constant $A_2^+$ is determined by the boundary conditions in Eqs. (21a)-(21c) after replacing $J_{0n} \to I_{0sn}$ and $\chi_n = 1$. After inserting Eqs. (39a) and (39b) into Eqs. (21b) and (21c), and performing the resulted integrals, both Eqs. (21b) and (21c) reduce to Eq. (34b). This is expected given that Eq. (34b) or its form in Eq. (A17) has been derived from the definition of the heat flux, which already takes into account the energy conservation principle, as demonstrated in the Appendix. Thus, it is clear that Eqs. (39a) and (39b) already take into account the constant heat flux condition at the steady-state, which represents another advantage of our analytical formalism. The value of $A_2^+$ is then given by the combination of Eqs. (21a) and (34b), which yields

$$A_2^+ = \frac{t_{12}}{r_{12}}\left(A_1^+ - \frac{3q_0}{4}(\lambda + 2\beta)\right). \tag{40}$$

The remaining constant $A_1^+$ is set by the specification of the temperature at the other side ($x \to \infty$) of the semi-infinite layer. Given that this value is usually not known explicitly, it is convenient to replace this condition by the temperature $T_0$ at the illuminated surface ($x = 0$) of the finite layer. Therefore $A_1^+ = \rho_1 c_1 v_1 T_0 / 4$ and Eqs. (39a) and (39b) can be written in terms of the temperature as

$$T_0 - T_1(x) = \frac{3q_0}{\rho_1 c_1 v_1}\left[x + \beta + \gamma\left(p(x) - p(\lambda - x)\right)\right], \tag{41a}$$

$$T_0 - T_2(x) = \frac{3q_0}{\rho_1 c_1 v_1}\left[\delta_{12}(x + p(x)) + \lambda + 2\beta\right], \tag{41b}$$

where we used the fact that $\delta_{12} = \rho_1 c_1 v_1 / \rho_2 c_2 v_2 = r_{12}/t_{12}$, as established by Eq. (19). Equations (41a) and (41b) describe fully the temperature inside the finite and semi-infinite layers shown in Fig. 2(a), by taking into account the effects of the film thickness and of the thermal mismatch between the layers $\rho_1 c_1 v_1 / \rho_2 c_2 v_2$.

## B. Modulated heat conduction

In this subsection, we are going to solve Eqs. (15) and (17) for the modulated heat conduction inside the finite and semi-infinite layers shown in Fig. 2(a). For simplicity, we are going to start with the semi-infinite layer in absence of the finite one, as we did for the steady-state problem.

- **Semi-infinite layer**

Based on Eqs. (15b) and (17b), the equilibrium intensity and heat flux inside the semi-infinite layer can be normalized as

$$2V_2(x) = E_2(\chi_2 x) + \int_0^\infty V_2(x')E_1(\chi_2|x-x'|)dx', \tag{42a}$$

$$\frac{\chi_2 q_{t2}(x)}{2A_2^+} = \chi_2 E_3(\chi_2 x) + \int_0^{\chi_2 x} V_2(x - y/\chi_2)E_2(y)dy - \int_0^\infty V_2(x + y/\chi_2)E_2(y)dy, \tag{42b}$$

where $V_2(x) = \pi J_{02}(x)/A_2^+$. In the diffusive limit ($x \to \infty$), the solution of the Fourier's law indicates an exponential decay for the temperature $V_2(x) = \chi_2 a \exp(-\chi_2 bx)$ and the heat flux $q_{t2}(x) = q_0 \exp(-\chi_2 bx)$, where $a$ and $b$ are constants. By direct comparison of Eq. (42) with the exact solution provided by the Fourier's law,[2] and using the relation $k = \rho c v l/3$ for the thermal conductivity, we find that $\chi_2 b = \sqrt{3i\omega\tau_2} = \sqrt{3(\chi_2 - 1)}$. Given that $\omega\tau_2 << 1$ is valid for most cases of practical interest, we are going to consider an approximation up to $\sqrt{i\omega\tau_2}$, which implies that $\chi_2 = 1 + i\omega\tau_2 \approx 1$ and $b = \sqrt{3i\omega\tau_2}$. The insertion of these results for $V_2$ and $q_{t2}$ into Eqs. (42a) and (42b) yields $a = 1$ and $A_2^+ = 3q_0/4b$. Similar to the steady state problem, the general solution of Eq. (42a) can then be written as the sum of the diffusive and ballistic contributions,

$$V_2(x) = e^{-bx} - bM(x), \tag{43}$$

where the function $M$ contains the non-Fourier effects, such that $M(x) \to 0$ for $x \to \infty$. By inserting Eq. (43) into Eq. (40a), it is found that under the present linear approximation in $b$, the function $M(x)$ satisfies an equation similar to Eq. (26a), for the function $g(x)$ involved in the steady state problem. This fact along with Eq. (26b) indicate that an accurate approximate solution of $M$ is given by

$$M(x) = A_0 E_2(x) + B_0 E_3(x). \tag{44}$$

where the constants $A_0$ and $B_0$ can be calculated by evaluating Eqs. (42a) and (42b) at $x = 0$:

$$2V_2(0) = 1 + \int_0^\infty V_2(y) E_1(y) dy, \tag{45a}$$

$$\frac{q_0}{A_2^+} = 1 - 2\int_0^\infty V_2(y) E_2(y) dy. \tag{45b}$$

The combination of Eqs. (43) - (45) yields the following system of equations

$$\begin{pmatrix} 2 - I_{12} & 1 - I_{13} \\ I_{22} & I_{23} \end{pmatrix} \begin{pmatrix} A_0 \\ B_0 \end{pmatrix} = \begin{pmatrix} 1/2 \\ 1/3 \end{pmatrix}, \tag{46}$$

where the coefficients $I_{nm}$ are defined in Eq. (31).

The solution of Eq. (46) for five decimal figures is $A_0 = -3.97433$ and $B_0 = 9.17085$. Thus, the solution of Eq. (42a) in terms of $J_{02}(x) = A_2^+ V_2(x)/\pi$ is

$$J_{02}(x) = \frac{3q_0}{4\pi} \left( \frac{e^{-bx}}{b} - M(x) \right), \tag{47a}$$

and the corresponding modulated temperature (see Eq. (6)) and heat flux (see Eq. (42b)) are

$$\psi_2(x) = \frac{3q_0}{\rho_2 c_2 v_2} \left( \frac{e^{-bx}}{b} - M(x) \right), \tag{47b}$$

$$q_{t2}(x) = q_0 \left[ e^{-bx} - b^2 M_1(x) \right], \tag{47c}$$

respectively. The function $M_1$ is defined in Eq. (A32) of the Appendix and shown in Fig. 3. The first term on the right-hand side of Eq. (47b) is the prediction of the Fourier's law while the function $M(x)$ represents the correction term due to the ballistic effect. At the illuminated surface $z = 0$, the function $M$ takes the value of $M(0) = 7/(17 - 8\ln(2)) = 0.61109$, and is bounded within the interval $0 \leq M(x) \leq 0.958$, as shown in Fig. 3. For a distance from the illuminated surface longer than the MFP, i.e., $x > 1$, the functions $E_n(x)$ and $M(x)$ decay faster than the exponential function, and therefore Eq. (47b) reduces to the prediction of the Fourier's law, as expected. Furthermore, given that $M_1(x) \leq M_1(0) = 1.070$, Eq. (47c) indicates that the non-Fourier effects on the heat flux inside a semi-infinite layer are negligible for low frequency when $b^2 = 3i\omega\tau_2 \to 0$,. Furthermore, we notice that the diffusion length defined under the formalism of the Fourier's law appears naturally in Eqs. (47a)-(47c) as $\mu_D = \sqrt{2}l_2/\sqrt{3\omega\tau_2}$, which coincides with its usual value provided that the thermal conductivity $k = \rho c v l/3$.

- **Thin Film Layer**

According to Eqs, (A19) and (A31) in the Appendix, the modulated equilibrium intensity $J_{01}$ and modulated heat flux $q_{t1}$ inside the finite layer shown in Fig. 2(a) are given by

$$J_{01}(x) = \frac{3q_0}{4\pi\eta_1}\left[Ae^{\eta_1 x} + Be^{-\eta_1 x} + C\big(M(x) - M(\lambda - x)\big)\right], \tag{48a}$$

$$q_{t1}(x) = q_0\left[-Ae^{\eta_1 x} + Be^{-\eta_1 x} + C\eta_1\big(M_1(x) + M_1(\lambda - x)\big)\right], \tag{48b}$$

$$A(1 + e^{\eta_1 \lambda}) + B(1 + e^{-\eta_1 \lambda}) = a_1^+ + a_1^-, \tag{48c}$$

where $a_1^{\pm} = (4\eta_1/3q_0)A_1^{\pm}$ and the functions $M$ and $M_1$ are defined in Eqs. (44) and (A32), respectively. The constants $A$, $B$, and $C$ can be determined in terms of $a_1^{\pm}$ by evaluating Eqs. (15a) and (15b) at $x = 0$:

$$2J_{01}(0) = \frac{A_1^+}{\pi} + \frac{A_1^-}{\pi}E_2(\lambda) + \int_0^\lambda J_{01}(y)E_1(y)dy, \tag{49a}$$

$$\frac{q_{t1}(0)}{2\pi} = \frac{A_1^+}{2\pi} - \frac{A_1^-}{\pi}E_3(\lambda) - \int_0^\lambda J_{01}(y)E_2(y)dy. \tag{49b}$$

Inserting Eq. (48a) into Eqs. (49), we find the following system of equations

$$\begin{pmatrix} 1+e^{\eta_1 \lambda} & 1+e^{-\eta_1 \lambda} & 0 \\ 2-\alpha_1^+ & 2-\alpha_1^- & \gamma \\ \alpha_2^+ - \dfrac{2}{3}\eta_1 & \alpha_2^- + \dfrac{2}{3}\eta_1 & \beta_2 \end{pmatrix} \begin{pmatrix} A \\ B \\ C \end{pmatrix} = \begin{pmatrix} 1 \\ 1 \\ \dfrac{1}{2} \end{pmatrix} a_1^+ + \begin{pmatrix} 1 \\ E_2(\lambda) \\ -E_3(\lambda) \end{pmatrix} a_1^-, \qquad (50)$$

where $\gamma = 2(M(0) - M(\lambda)) - \beta_1$ and

$$\alpha_n^\pm = \int_0^\lambda e^{\pm \eta_1 y} E_n(y)\, dy, \qquad (51a)$$

$$\beta_n = \int_0^\lambda \bigl(M(y) - M(\lambda - y)\bigr) E_n(y)\, dy. \qquad (51b)$$

Equation (50) describes fully the size- and the frequency- dependence of the parameters $A$, $B$, and $C$ for a layer with normalized thickness $\lambda = L/l_1$. For a semi-infinite layer ($\lambda \to \infty$, $a_1^- \to 0$), the solution of Eq. (50) is $(A,B,C) = a_1^+(0,1,-\eta_1)$, which reduces Eq. (48a) to Eq. (47a) for the semi-infinite layer, as should be. Furthermore, assuming that the surfaces $x = 0, \lambda$ of the finite layer are kept at temperatures $T_1$ and $T_2(<T_1)$, respectively, the constants $A_1^\pm$ are then given by $A_1^+ = \sigma_1 T_1^4$ and $A_1^- = \sigma_1 T_2^4$ (see Eq. (5)), which determine $a_1^\pm = (4\eta_1/3q_0) A_1^\pm$.

- **Two-Layer System**

The modulated components of the temperature and heat flux inside the two-layer system shown in Fig. 2(a) are derived in this subsection. These results are then compared with those predicted by the Fourier's law for the diffusive regime. According to our previous analysis for a thin film layer in the semi-infinite limit, the modulated equilibrium intensity $J_{02}$ and modulated heat flux $q_{t2}$ inside the semi-infinite layer in thermal contact with the finite layer shown in Fig. 2(a) can be written as follows

$$J_{02}(x) = \frac{3q_0}{4\pi\eta_2} a_2^+ \left( e^{-\eta_2 x} - \eta_2 M(x) \right), \qquad (52a)$$

$$q_{t2}(x) = q_0 a_2^+ \left[ e^{-\eta_2 x} - \eta_2^2 M_1(x) \right]. \qquad (52b)$$

where $a_2^+ = (4\eta_2/3q_0)A_2^+$. After combining Eqs. (48a) and (52a) with the boundary conditions in Eqs. (21a)-(21c), the following system of equations is obtained for the constants $a_1^\pm$

$$\begin{pmatrix} D^+ + E_3(\lambda) & \zeta \\ d^+ - 1/2 & d^- + E_3(\lambda) \end{pmatrix} \begin{pmatrix} a_1^+ \\ a_1^- \end{pmatrix} = -\frac{2\eta_1}{3}\begin{pmatrix} 0 \\ 1 \end{pmatrix}, \tag{53}$$

where we have used the fact that $a_2^+ = \sqrt{\tau_2/\tau_1}\, a_1^-/\delta_{12}$ (Eq. (21a)),

$$\zeta = D^- - \frac{1}{\delta_{12}}\left(\frac{1}{2t_{12}} + \frac{\eta_2}{3} + \frac{\ln(1+\eta_2)-\eta_2}{\eta_2^2}\right), \tag{54a}$$

$$d^\pm = \alpha_2^+ a^\pm + \alpha_2^- b^\pm + \beta_2 c^\pm, \tag{54a}$$

$$D^\pm = \alpha_2^- e^{\eta_1\lambda} a^\pm + \alpha_2^+ e^{-\eta_1\lambda} b^\pm - \beta_2 c^\pm, \tag{54a}$$

and the parameters $a^\pm$, $b^\pm$, and $c^\pm$ are given by Eq. (50), such that $A = a^+ a_1^+ + a^- a_1^-$, $B = b^+ a_1^+ + b^- a_1^-$, and $C = c^+ a_1^+ + c^- a_1^-$. Equations (50) and (54) fully determine the coefficients involved in the modulated equilibrium intensities and modulated heat fluxes inside the finite and semi-infinite layers shown in Fig. 2(a). The analytical expressions in Eqs. (48) and (52) contain the effects of the layer thickness, modulation frequency, phonon mean free paths, phonon relaxation times, and thermal mismatch between the layers through the ratios $\delta_{12} = \rho_1 c_1 v_1/\rho_2 c_2 v_2$ and $\tau_2/\tau_1$. Finally the modulated temperature profiles are given by Eq. (6), where $\psi_n(x) = 4\pi J_{0n}(x)/\rho_n c_n v_n$.

### III. RESULTS AND DISCUSSIONS

In this section, both the steady state and modulated temperature profiles are analyzed as a function of the film thickness and the modulation frequency. For the two-layer system shown in Fig. 2(a), calculations were done with the data reported in Table I.

### A. Steady-state temperature profiles

Figures 5(a), 5(b) and 5(c) show the steady-state temperature distribution inside a semi-infinite layer (Eq. (33)), a layer with a finite size (Eq. (34a)), and a two-layer system (Eqs. (41a) and (41b)), respectively. As expected, when the distance from the excited surface $z = 0$ of the semi-infinite layer increases to values much longer than the phonon MFP, the temperature predicted by the BTE reduces to the straight line as predicted by the Fourier's law. By contrast, as this distance takes any smaller values than a phonon MFP, the temperature determined by BTE tends to the constant value $T_B = T_0 - \sqrt{3}q_0/\rho cv$, which differs from the predictions of the Fourier's law. This is reasonable given that the emitted phonons at $z = 0$ travel ballistically within an average MFP distance, and therefore the temperature does not change remarkable within a phonon MFP distance.[1,6] The fact that $T_0 > T_B$ indicates that the temperature imposed at the outer surface $z = 0^+$ is higher than the emitted phonons at the inner surface $z = 0^-$.

Similar to that in the semi-infinite medium in Fig. 5(a), a temperature jump is also observed in Fig. 5(b) at the boundaries of a finite layer. These jumps increase as the layer thickness reduces. This is reasonable considering that the "hot" phonons emitted from the surface $z = 0$ heat up the "cold" phonons emitted from the opposite surface $z = L$ and vice versa. In the extreme case of ballistic limit ($L \ll l_1$), the temperature inside the layer is given by $(T_1 + T_2)/2$. It is clear from Fig. 5(b) that the analytical results of the present work are in very good agreement with the dotted lines obtained through numerical simulations of phonon Boltzmann transport[1,6] The advantage of using the analytical approach over the numerical one is the simple description of the temperature through Eq. (34a), which is able to provide physical insights without much of numerical efforts. As the layer thickness is scaled down to values comparable to the phonon MFP, the temperature tends to become independent of the position, which is the signature of the ballistic heat conduction. The temperature profiles and the jumps at the interfaces are very similar to those observed for photon radiative transfer in a plane-parallel medium.[31] The combined behavior of the temperature inside the semi-infinite and finite layers arises in the two-layer system, as shown in Fig. 5(c).

**Table 1.** Room-temperature material properties used in the calculations.[8]

| Material | Bulk thermal conductivity (W/m.K) | MFP (nm) | Debye temperature (K) | Average phonon group velocity (km/s) | Analogous Stephan-Boltzmann constant (W/m$^2$K$^4$) |
|---|---|---|---|---|---|
| Si | 150 | 268 | 645 | 8.433 | 107.49 |
| Ge | 51.7 | 171 | 374 | 5.400 | 262.15 |

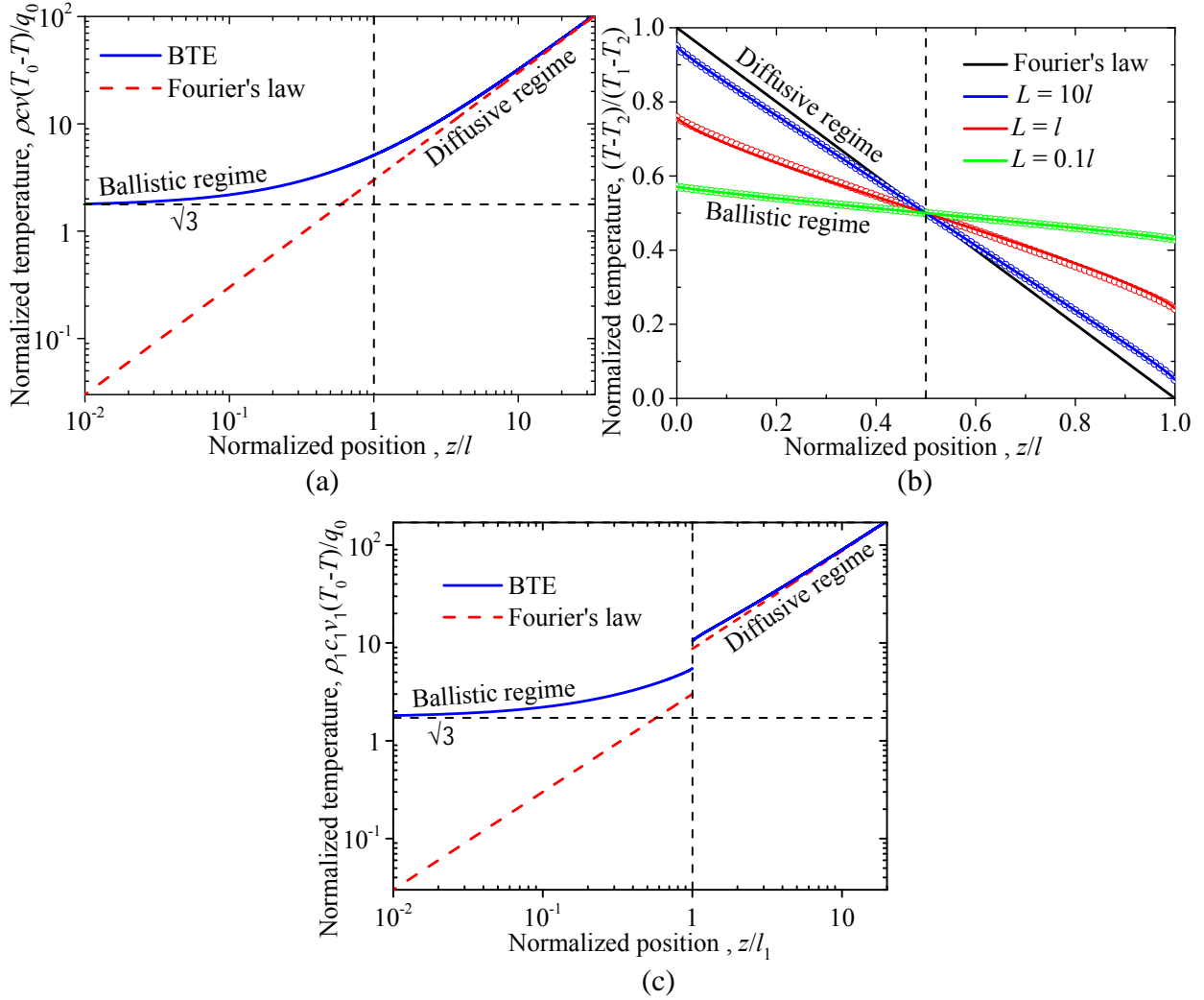

*FIG. 5. Steady-state temperature profile inside: (a) a semi-infinite layer, (b) a layer with finite thickness, and (c) the two-layer system shown in Fig. 1(b). Dotted lines in (b) correspond to the numerical predictions reported by Chen,[6] and the calculations in (c) were performed with the data of Table I for a film of Si of thickness $L = l_1$ deposited on a semi-infinite Ge substrate. The predictions of the Fourier's law in Fig. 5(c) were calculated using the interface thermal resistance $R = 2\left(1/\rho_1 c_1 v_1 + 1/\rho_2 c_2 v_2\right)$, which is appropriate for the diffusive interface scattering.[5]*

The cross-plane thermal conductivity defined in Eq. (38) for a single layer is shown in Fig. 6, as a function of its normalized thickness. This result agrees well with the work by Majumdar[1] in the pure ballistic ($L \ll l_1$) regime, but they differ in the intermediate diffusive-ballistic regime. This is because the parameter $\beta$ involved in Eq. (38) is different than its ballistic value 2/3, within this thickness range, as shown in Fig. (4). Given that our formula for the thermal conductivity has been derived rigorously from the analytical solution of the phonon BTE, its predictions are expected to be an extension of Majumdar model, which was inferred from numerical simulations.[1]

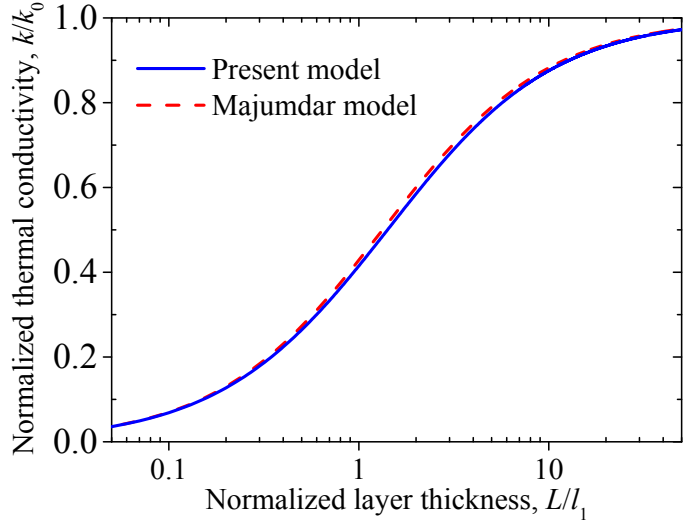

*FIG. 6. Comparison of the cross-plane thermal conductivity of a layer predicted by the present and Majumdar[1] models, as a function of its normalized thickness.*

## B. Modulated temperature profiles

The normalized amplitude and phase of the modulated temperature at the illuminated surface $z = 0$ of a semi-infinite layer are shown in Figs. 7(a) and 7(b), respectively. We have considered the frequency interval $\omega\tau \leq 0.1$, because the analytical result in Eq. (47b) is only valid for $\omega\tau \ll 1$. In the low frequency regime ($\omega\tau \to 0$), the amplitudes and phases predicted by the BTE and Fourier's law agree with each other, as expected. Given that the average relaxation time of phonons in most materials is shorter than hundreds of picoseconds ($\tau < 10^{-10}$ s),[8] this fact indicates that the Fourier's law holds when the frequency $f = \omega/2\pi \ll 1$ GHz. On the other hand, for frequencies close to but smaller than $\omega = 0.1\tau^{-1}$, the predictions of the BTE and Fourier's law differ, especially on the phase signal. As the frequency increases, the phase predicted by the Fourier's law remains independent of the frequency, while the one predicted by the BTE decreases. Taking into account that the BTE is much more general than the Fourier's law, the modulated heat conduction at high frequency ($f > 1$ GHz) is expected to be better described by the BTE.

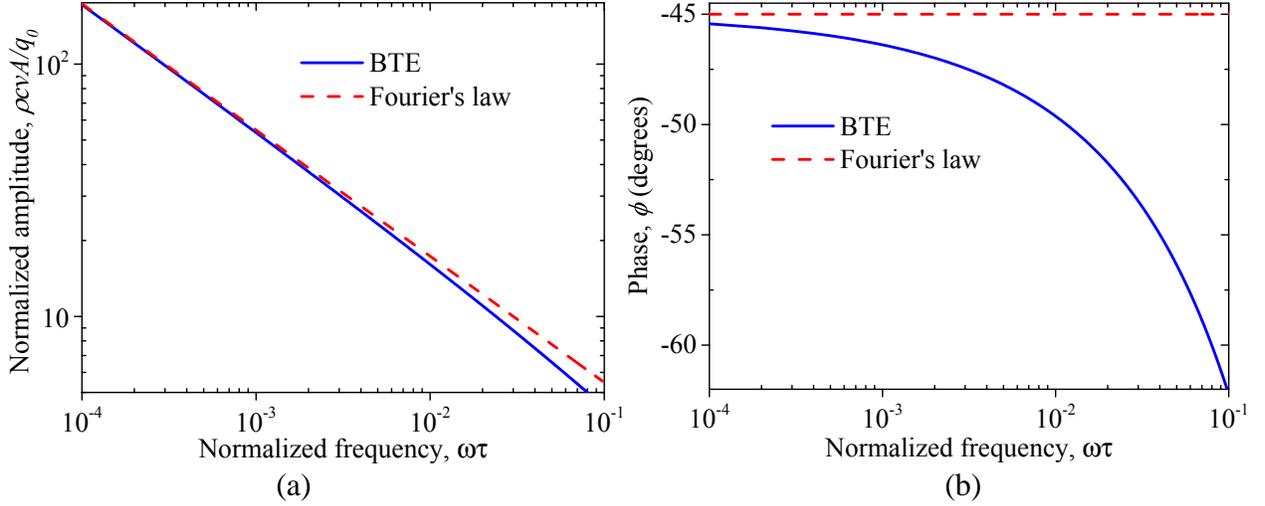

*FIG. 7. Frequency dependence of the (a) normalized amplitude and (b) phase of the modulated temperature at the illuminated surface $z = 0$ of a semi-infinite layer.*

Figures 8(a) and 8(b) shows the normalized amplitude and phase of the modulated temperature profile inside a semi-infinite layer as a function of its position. Both of these signals exhibit a similar behavior with the position and they become closer to the diffusive limit determined by the Fourier's law, when the position increases to values larger than the phonon MFP. The slight difference of the BTE and Fourier's law predictions for $z \leq l$ follows a similar behavior as the one presented in the corresponding state-state components shown in Fig. 5(a) and it is due to the ballistic behavior of phonons traveling within this position range. The relatively lower values predicted by the BTE (less than 0.5 degrees in the phase) could be associated with the continuous emission of phonons from $z = 0$ without receiving feedback of phonons from $z \to \infty$, which create a zone of ballistic phonons ($0 < z \leq l$), whose modulated temperature is slightly smaller than the one they would have in equilibrium.

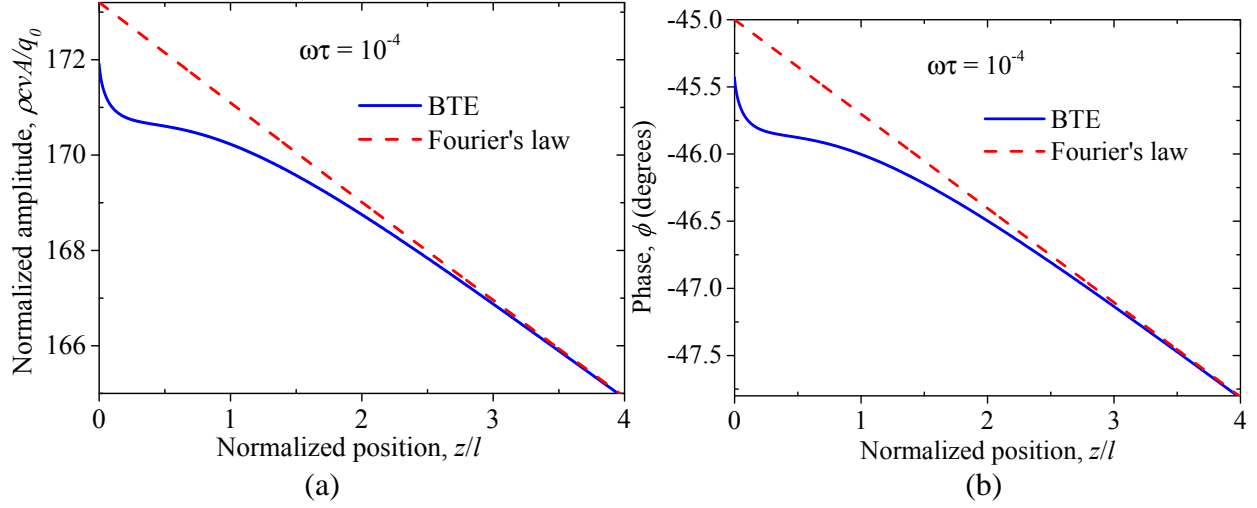

*FIG. 8. Position dependence of the (a) normalized amplitude and (b) phase of the modulated temperature inside a semi-infinite layer.*

The frequency dependence of the normalized amplitude and phase of the modulated temperature at the illuminated surface of a Si film deposited on a semi-infinite Ge substrate are shown in Figs. 9(a) and 9(b), respectively. In the limit case of a thick film ($L = 100 l_1$) and very low modulation frequency ($\omega \tau_1 \to 0$), both the amplitude and phase reduces to its corresponding values predicted by the Fourier's law, as should be. However, for a thinner film and/or higher frequencies, these signals take lower values than those predicted by the Fourier's law, such that their difference increases as the film thickness reduces and/or the frequency increases. A similar behavior is shown in Figs. 10(a) and 10(b) for the position dependence of the normalized amplitude and phase inside the layers. This is consistent with the observation shown in Fig. 7 and 8 for a semi-infinite layer and it indicates the failure of the Fourier's law to capture the ballistic features of the ultrafast modulated heat conduction in very thin layers, as discussed by Joseph and Preziosi,[32] and the references therein.

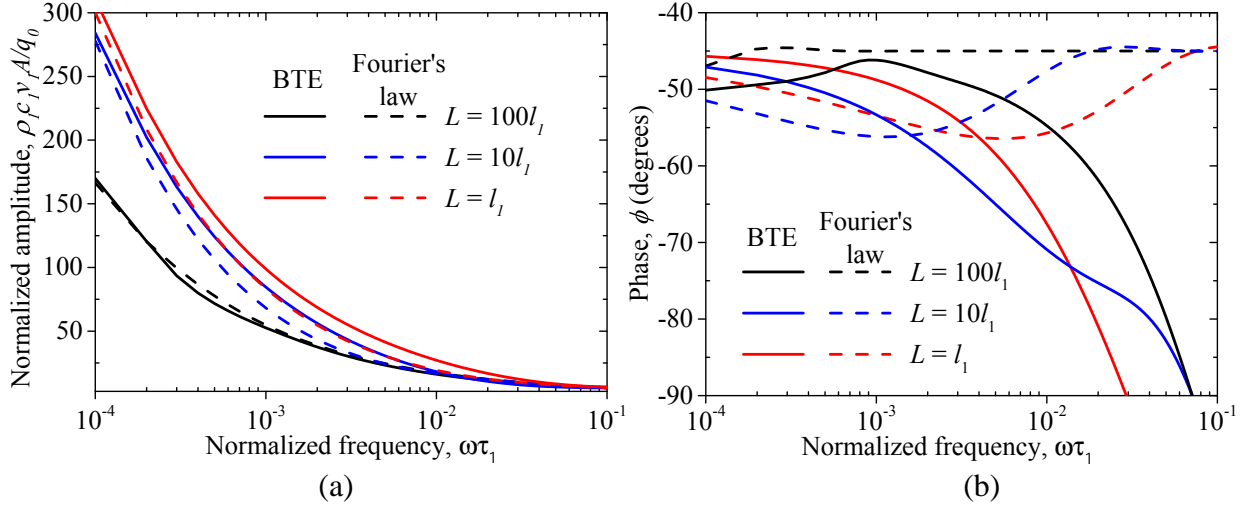

*FIG. 9. Frequency dependence of the (a) normalized amplitude and (b) phase of the modulated temperature at the illuminated surface $z_1 = 0$ of a finite Si layer deposited on a semi-infinite Ge substrate. Calculations were performed with the data reported in Table I. The predictions of the Fourier's law were calculated using the interface thermal resistance $R = 2\left(1/\rho_1 c_1 v_1 + 1/\rho_2 c_2 v_2\right)$, which is appropriate for the diffusive regime.[5]*

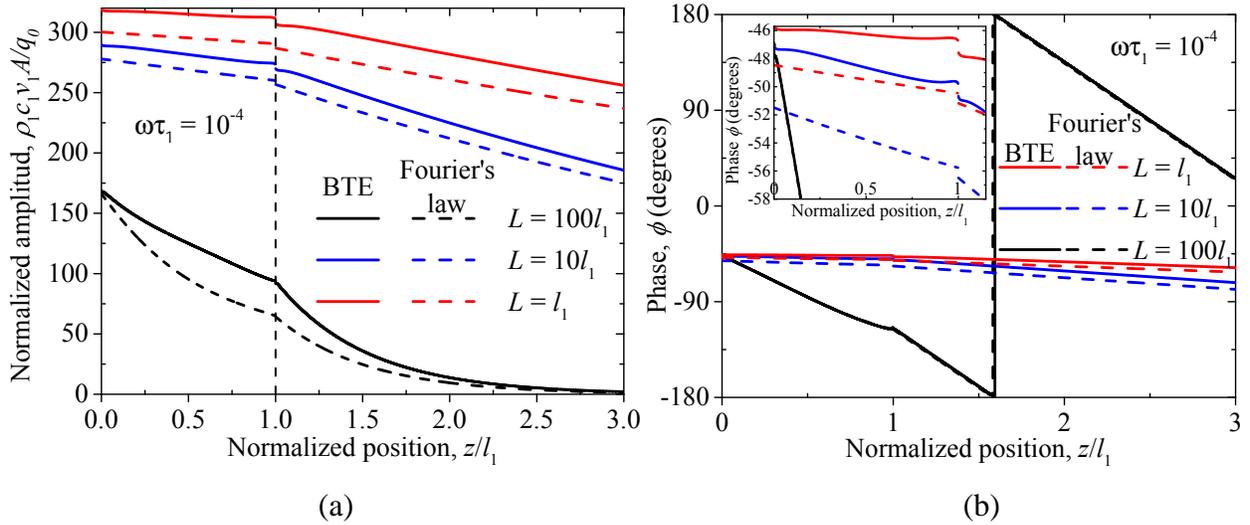

*FIG. 10. Position dependence of the (a) normalized amplitude and (b) phase of the modulated temperature inside a finite Si layer deposited on a semi-infinite Ge substrate. Calculations were performed with the data reported in Table I.*

Figures 11(a) and 11(b) show the position dependence of the normalized amplitude and phase of the modulated heat flux inside a finite Si film deposited on a semi-infinite Ge substrate, respectively. Regardless of a particular film thickness or an inner position, the heat fluxes predicted by the BTE agree quite well with the corresponding predictions of the Fourier's law, such that the amplitude increases as the film thickness is scaled down. This behavior holds for any frequency when $\omega\tau_1 < 0.1$. This fact establishes that the propagation of the modulated heat flux inside the layers is not significantly affected by the ballistic and high frequency effects observed on the modulated temperature, as shown in Figs. 9 and 10. Therefore even though the energy flux propagates essentially as described by the Fourier's law, the modulated temperature is more properly described by the BTE.

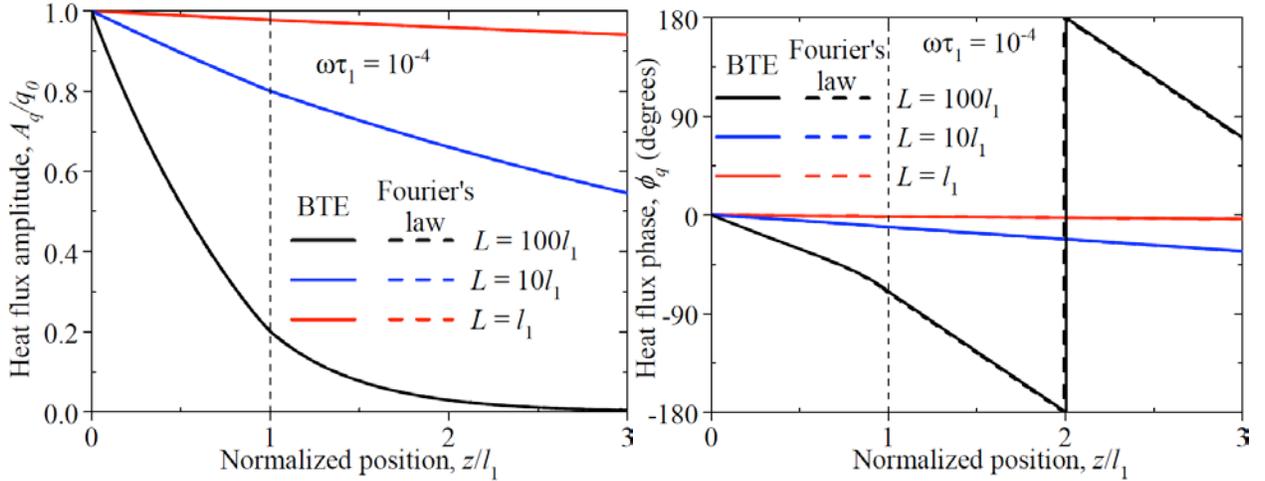

*FIG. 11. Position dependence of the (a) normalized amplitude and (b) phase of the modulated heat flux inside a Si layer in thermal contact with a semi-infinite one of Ge. Calculations were performed with the data reported in Table I.*

## IV. CONCLUSIONS

Steady state and modulated components of the temperature inside a dielectric film deposited on a semi-infinite substrate have been analyzed by solving analytically the phonon Boltzmann transport equation under the relaxation time approximation. It has been shown that both the steady state and modulated components of temperature depend strongly on the ratio between the film thickness and the phonon mean free path, and they reduces to their corresponding diffusive values predicted by the Fourier's law as this ratio increases. By contrast, in the ballistic regime in which this ratio is comparable to the unity, the steady state temperature tends to be independent of position, and the amplitude and phase of the modulated temperature display lower values than those determined by the Fourier's law, such that their corresponding difference increases with modulation frequencies higher than 1 MHz. Furthermore, an invariant of heat conduction and a simple formula for the cross-plane thermal conductivity of dielectric thin films are obtained, which could be useful for understanding and optimizing their thermal performance. This work could serve as the theoretical basis for the determination of phonon mean free path using transient thermoreflectance method.

**Acknowledgments**: This work is supported by the NSF CAREER award (Grant No. 0846561) and AFOSR Thermal Sciences Grant (FA9550-11-1-0109).

**Appendix: Analytical Solution of the BTE Obtained using the Discrete-Ordinates Method**

To obtain the solution of the phonon BTE, we are going to use the discrete-ordinates method,[33] which has widely been used to solve numerically the Boltzmann transport of photons, neutrons and most recently phonons. In this appendix, we are going to show that this method can also provide analytical solutions for both the steady state and modulated components of the phonon temperature. This method is based on the Gaussian quadrature

$$\int_{-1}^{1} F(\mu) d\mu = \sum_{n} a_n F(\mu_n), \tag{A1}$$

where the coefficients $a_n = a_{-n}$ are known for $n = \pm 1, \pm 2, \ldots \pm N$. Thus, the quadrature method consists of splitting the interval $-1 \leq \mu \leq 1$ to $2N$ symmetrical directions ($\mu_{-n} = -\mu_n$). For $F(\mu) = \mu^i$, Eq. (A1) yields

$$\sum_{n} a_n \mu_n^i = \begin{cases} \dfrac{2}{i+1}, & i = 0, 2, 4, \ldots \\ 0, & i = 1, 3, 5, \ldots \end{cases} \tag{A2}$$

- **Steady-state heat conduction problem**

According to Eqs. (10a), (13) and (A1), the steady-state BTE for the phonon intensity $I_s$ is given by

$$I_s(x,\mu) + \mu \frac{\partial I_s(x,\mu)}{\partial x} = \frac{1}{2} \int_{-1}^{1} I_s(x,\mu) d\mu = \frac{1}{2} \sum_{n} a_n I_s(x,\mu_n), \tag{A3}$$

which represents a system of $2N$ linear differential equations for any direction $\mu = \mu_m$ ($m = \pm 1, \pm 2, \ldots \pm N$). It can be shown that the solutions of Eq. (A3) have the exponential form $I_s(x,\mu_n) = C_n \exp(-\delta x)$, where $C_n$ and $\delta$ are constants. Substituting this expression into Eq. (A3) yields

$$C_m = \frac{D}{1 - \mu_m \delta}, \tag{A4}$$

where $2D = \sum_{n} a_n C_n$ is a constant independent of $n$. Inserting again the exponential solution into Eq. (A3) and using Eq. (A4), the values of the exponent $\delta$ are determined by the following characteristic equation,

$$\sum_n \frac{a_n}{1-\mu_n \delta} = 2. \tag{A5}$$

According to Eq. (A2) for $i = 1$, Eq. (A5) can be conveniently rewritten as

$$\sum_n \frac{a_n \mu_n}{1-\mu_n \delta} = \sum_{n=1}^{N} \frac{a_n \mu_n^2}{1-(\mu_n \delta)^2} = 0, \tag{A6}$$

which is polynomial of degree $N-1$ in $\delta^2$. Therefore, the $2(N-1)$ roots of Eq. (A6) have the form $\delta = \pm \delta_j$ for $j = 1,2,...,N-1$. Thus, the first $2(N-1)$ solutions of the system in Eq. (A3) are determined by

$$I_s(x,\mu_n) = \sum_{j=1}^{N-1} \left( \frac{D_j e^{-\delta_j x}}{1-\mu_n \delta_j} + \frac{D_{-j} e^{\delta_j x}}{1+\mu_n \delta_j} \right), \tag{A7}$$

where $D_j$ and $D_{-j}$ are arbitrary constants. Given that the system of equations in Eq. (A3) is of order $2N$, we require two more independent solutions. Note that Eqs. (A2) and (A5) can be written as $\sum_{n=1}^{N} a_n = 1 = \sum_{n=1}^{N} \frac{a_n}{1-(\mu_n \delta)^2}$, which clearly indicate that $\delta = 0$ is a double root of the characteristic equation. Based on the theory of differential equations, the intensity corresponding to this double root is given by $I_s(x,\mu_n) = \alpha'_n (x + \beta'_n)$, where $\alpha'_n$ and $\beta'_n$ are constants. The substitution of this expression into Eq. (A3) yields $\alpha'_m = \sum_n a_n \alpha'_n / 2 = \alpha'$ and $\beta'_m + \mu_m = \sum_n a_n \beta'_n / 2 = \beta'$. The general solution of the system in Eq. (A3) can therefore written as

$$I_s(x,\mu_n) = \alpha' \left[ x + \beta' - \mu_n + \sum_{j=1}^{N-1} \left( \frac{D_j e^{-\delta_j x}}{1-\mu_n \delta_j} + \frac{D_{-j} e^{\delta_j x}}{1+\mu_n \delta_j} \right) \right], \tag{A8}$$

where the constants $D_j$ and $D_{-j}$ appearing in Eq. (A7) have been redefined to absorb $\alpha'$. Note that Eq. (A8) is valid for any $-1 \leq \mu_n \leq 1$ and therefore it can also be written with the substitution $\mu_n \to \mu$. Equation (A8) shows clearly that the steady-state intensity has both diffusive and ballistic contributions provided by its linear and exponential dependence on the position, respectively. For phonon heat conduction across a layer with diffuse scattering at its boundaries, Eqs. (11) and (12) establishes that for $0 < \mu \leq 1$, the boundary conditions of $I_s$ are given by

$$I_s(0,\mu) = A^+/\pi, \tag{A9}$$

$$I_s(\lambda,-\mu) = A^-/\pi, \tag{A10}$$

where $\lambda = L/l$ is the ratio between the layer thickness and the MFP of phonons inside the medium. After inserting Eq. (A8) into Eqs. (A9) and (A10) and summing the obtained results, the following is obtained

$$\alpha'(\lambda + 2\beta') = \frac{A^+ + A^-}{\pi}, \tag{A11}$$

$$D_{-j} = -D_j e^{-\delta_j \lambda}. \tag{A12}$$

The intensity $I_s$ is now expressed as

$$I_s(x,\mu_n) = \alpha'\left[x + \beta' - \mu_n + \sum_{j=1}^{N-1} D_j\left(\frac{e^{-\delta_j x}}{1-\mu_n \delta_j} - \frac{e^{-\delta_j(\lambda-x)}}{1+\mu_n \delta_j}\right)\right]. \tag{A13}$$

Based on Eqs. (7), (13) and (A1), the equilibrium intensity $I_{0s}$ (temperature) and the heat flux $q_s$ are given by

$$2I_{0s}(x) = \int_{-1}^{1} I_s(x,\mu)d\mu = \sum_n a_n I_s(x,\mu_n), \tag{A14}$$

$$q_s(x) = 2\pi \int_{-1}^{1} I_s(x,\mu)\mu d\mu = 2\pi \sum_n a_n \mu_n I_s(x,\mu_n). \tag{A15}$$

By inserting Eq. (A13) into Eqs. (A14) and (A15), and evaluating the involved sums with Eqs. (A2), (A5) and (A6), the following results are obtained

$$I_{0s}(x) = \alpha'\left[x + \beta' + \sum_{j=1}^{N-1} D_j\left(e^{-\delta_j x} - e^{-\delta_j(\lambda-x)}\right)\right], \tag{A16}$$

$$q_s = -\frac{4\pi}{3}\alpha'. \tag{A17}$$

Equation (A17) shows that the heat flux is a constant independent of the position, which is consistent with the principle of energy conservation for the steady-state heat conduction. Based on Eqs. (A8) and (A15), it can be shown that this result is independent of the diffuse boundary conditions in Eqs. (A9) and (A10), as should be. Note that the sum on each exponential term in Eq. (A16) can be considered as the series expansion of a given function and therefore they can be written as follows

$$I_{0s}(x) = \alpha'\left[x + \beta' + \gamma\left(p(x) - p(\lambda - x)\right)\right], \tag{A18}$$

where $\gamma$ is a constant. Equation (A18) can then be expressed in a more meaningful way in term of the normalized equilibrium intensity $U = \left(\pi I_{0s} - A^-\right)/\left(A^+ - A^-\right)$,

$$U(x) = 1 - \alpha\left[x + \beta + \gamma\left(p(x) - p(\lambda - x)\right)\right], \tag{A19}$$

$$\alpha = \frac{3}{4}Q_s = \frac{1}{\lambda + 2\beta}, \tag{A20}$$

where $Q_s = q_s/\left(A^+ - A^-\right)$. Equation (A19) represents the general solution for phonon temperature, in which the parameters $\beta = \beta(\lambda)$ and $\gamma = \gamma(\lambda)$ are determined by the boundary conditions on the temperature and heat flux.

By comparing Eq. (A19) with Eq. (28a) for a semi-infinite medium ($\lambda \to \infty$), it is clear that $\beta(\infty) = p(\infty)$, $\gamma(\infty) = 1$, and $p(x)$ is given by Eq. (28b). Based on Eqs. (A16) or (A18), the following relation

$$I_{0s}(x) + I_{0s}(\lambda - x) = 2I_{0s}(\lambda/2) = \frac{A^+ + A^-}{\pi}, \tag{A22}$$

is obtained to be valid for any position $0 \leq x \leq \lambda$. Equation (A22) establishes that the sum of the equilibrium intensities at two equidistant points from the external surfaces of a finite layer, is an invariant of heat conduction. For temperatures smaller than the Debye temperature, this indicates that if we impose the temperatures $T_1$ and $T_2$ at the external surfaces of a finite layer as shown in Fig. A1, then $A^+ = \pi\sigma T_1^4$, $A^- = \pi\sigma T_2^4$, and $T^4(x) + T^4(\lambda - x) = T_1^4 + T_2^4$. In particular, $T^4(\lambda/2) = (T_1^4 + T_2^4)/2$ is obtained at the center of the layer, which is the average of the fourth powers of the temperatures at the external surfaces. This feature of temperature is an exclusive prediction of the phonon BTE, which cannot be predicted by the Fourier's law. An analogous behavior for temperature was found in the radiative heat transfer.[30]

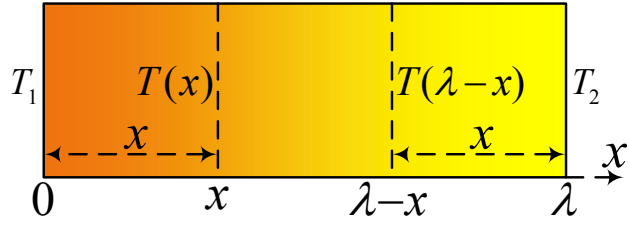

*FIG. A1*. *Temperature distribution inside a finite layer.*

- **Modulated heat conduction problem**

Similar to the steady-state case, we start with the numerical form of the phonon BTE for the modulated component of the temperature

$$\chi J(x,\mu) + \mu \frac{\partial J(x,\mu)}{\partial x} = \frac{1}{2}\int_{-1}^{1} J(x,\mu) d\mu = \frac{1}{2}\sum_n a_n J(x,\mu_n), \quad (A22)$$

The characteristic equation of Eq. (A22) can be derived by following a similar procedure as the one done for the steady-state problem. The final result is

$$\sum_n \frac{a_n}{\chi - \mu_n \sigma} = 2\sum_{n=1}^{N} \frac{\chi a_n}{\chi^2 - (\mu_n \sigma)^2} = 2. \quad (A23)$$

With the help of Eq. (A2) for $i = 1$, Eq. (A23) can be rewritten in a form similar to Eq. (A6) as,

$$\sum_{n=1}^{N} \frac{\chi(\chi-1) - (\mu_n \delta)^2}{\chi^2 - (\mu_n \delta)^2} a_n = 0. \quad (A24)$$

As discussed earlier, we are only interested in the case of $\omega\tau \ll 1$ for practical applications. Hence for an approximation up to $\sqrt{3i\omega\tau} = \sqrt{3(\chi-1)}$, Eq. (A24) reduces to Eq. (A6), which has been derived for the steady-state case. The roots of Eq. (A24) are then real and given by $\delta = \pm\delta_j$ for $j = 1,2,...,N-1$. To find the two additional roots, note that Eq. (A24) suggests that there exist a constant $\mu_n = \mu_0$, such that $(\mu_0\delta)^2 = \chi(\chi-1)$. Note that in this case, the second equality in Eq. (A22) becomes $\int_{-1}^{1} J(x,\mu) d\mu = 2J(x,\mu_0)$, which is nothing more than the well-known mean value theorem of integrals, applied to the modulated intensity *J*. Therefore, the existence of the $\mu_0$ is supported by this theorem, the remaining two solutions are therefore $\delta\mu_0 = \pm\sqrt{\chi(\chi-1)} \approx \pm\sqrt{i\omega\tau}$, and the general solution for $J$ is

$$J(x,\mu_n) = Ae^{\eta x} + Be^{-\eta x} + \sum_{j=1}^{N-1}\left(\frac{C_j e^{-\delta_j x}}{\chi - \mu_n \delta_j} + \frac{C_{-j} e^{\delta_j x}}{\chi + \mu_n \delta_j}\right), \tag{A25}$$

where $\eta = \sqrt{i\omega\tau}/\mu_0$ and $A$, $B$, $C_j$ and $C_{-j}$ are arbitrary constants. By applying the boundary conditions of diffuse scattering for the modulated intensity $J$ (see Eqs. (A9) and (A10)), it can be shown that

$$A(1+e^{\eta\lambda}) + B(1+e^{-\eta\lambda}) = \frac{A^+ + A^-}{\pi}, \tag{A26}$$

$$C_{-j} = -C_j e^{-\delta_j \lambda}. \tag{A27}$$

The intensity $J$ can now be expressed as

$$J(x,\mu_n) = Ae^{\eta x} + Be^{-\eta x} + \sum_{j=1}^{N-1} C_j \left(\frac{e^{-\delta_j x}}{\chi - \mu_n \delta_j} - \frac{e^{-\delta_j(\lambda-x)}}{\chi + \mu_n \delta_j}\right), \tag{A28}$$

and the application of Eq. (A14) for the modulated problem yields

$$J_0(x) = Ae^{\eta x} + Be^{-\eta x} + \sum_{j=1}^{N-1} C_j \left(e^{-\delta_j x} - e^{-\delta_j(\lambda-x)}\right). \tag{A29}$$

The sum on each exponential term in Eq. (A29) represents the series expansion of a given function and therefore they can be written as follows

$$J_0(x) = Ae^{\eta x} + Be^{-\eta x} + C\big(M(x) - M(\lambda - x)\big), \tag{A30}$$

where $C$ is a constant independent of the position $x$. The comparison of Eq. (A30) with Eq. (41a) for a semi-infinite medium ($\lambda \to \infty$) determine that $\mu_0 = 1/\sqrt{3}$, $M(\infty) = 0$ and $M(x)$ is defined by Eq. (38). Equation (A30) is the general solution for the modulated intensity and therefore for the modulated temperature. Given the dependence between $A$ and $B$ in Eq. (A26), only two of the three constants in Eq. (A30) are independent. In contrast to the steady-state problem, it is important to note that Eq. (A28) predicts a modulated heat flux that depends on the position, which is consistent with Eqs. (17a) and (17b). Furthermore, the sum $J_0(x) + J_0(\lambda - x)$ is independent of the function $M$, such that $J_0(0) + J_0(\lambda) = (A^+ + A^-)/\pi$. This equality holds only at the external surfaces $x = 0, \lambda$ of the layer, and not at any position inside the layer like the case of the steady-state problem.

Finally, the combination of Eqs. (A15) and (A28) yields the following modulated heat flux

$$q_t(x) = \frac{4\pi}{3}\eta\left[-Ae^{\eta x} + Be^{-\eta x} + C\eta\left(M_1(x) + M_1(\lambda - x)\right)\right], \tag{A31}$$

$$M_1(x) = A_0 E_3(x) + B_0 E_4(x) \tag{A32}$$

where $M_1$ is the negative integral of $M$ defined in Eq. (38). Note that the derivative of $q_t$ is related to the equilibrium intensity in Eq. (A30) through the simple relation $q_t'(x) = -4\pi(\chi - 1)J_0(x)$. This result can also be derived using the integral Eqs. (15a) and (17a), which shows the consistence of Eqs. (A30) and (A31).